\documentclass[prc,showpacs,superscriptaddress]{revtex4} 
\usepackage{rotating}
\usepackage{graphicx}
\usepackage{wasysym}
\begin{document}

\title{Measurements of ultracold neutron upscattering and absorption\\ in polyethylene 
and vanadium.}

\author{E. I. Sharapov}
\affiliation{Joint Institute for Nuclear Research, 141980, Dubna, Russia}
\author{C. L. Morris}
\thanks{Corresponding author}
\email{morris@lanl.gov}
\affiliation{Los Alamos National Laboratory, Los Alamos, NM 87544, USA}
\author{M. Makela}
\affiliation{Los Alamos National Laboratory, Los Alamos, NM 87544, USA}
\author{A. Saunders}
\affiliation{Los Alamos National Laboratory, Los Alamos, NM 87544, USA}
\author{Evan R. Adamek}
\affiliation{Department of Physics,Indiana University, Indiana 47405-7105 USA}
\author{Y. Bagdasarova}
\affiliation{Los Alamos National Laboratory, Los Alamos, NM 87544, USA}
\author{L. J. Broussard}
\affiliation{Los Alamos National Laboratory, Los Alamos, NM 87544, USA}
\author{C. B. Cude-Woods}
\affiliation{Department of Physics,Indiana University, Indiana 47405-7105 USA}
\author{Deon E Fellers}
\affiliation{Los Alamos National Laboratory, Los Alamos, NM 87544, USA}
\author{Peter Geltenbort}
\affiliation{Institut Laue-Langevin, 38042 Grenoble Cedex 9, France}
\author{S. I. Hasan}
\affiliation{Department of Physics and Astronomy, University of Kentucky, Lexington, Kentucky 40506, USA}
\author{K. P. Hickerson}
\affiliation{Kellogg Radiation Laboratory, California Institute of Technology, Pasadena, California 91125, USA.}
\author{G. Hogan}
\affiliation{Los Alamos National Laboratory, Los Alamos, NM 87544, USA}
\author{A. T. Holley}
\affiliation{Department of Physics,Indiana University, Indiana 47405-7105 USA}
\author{Chen-Yu Liu}
\affiliation{Department of Physics,Indiana University, Indiana 47405-7105 USA}
\author{M. P. Mendenhall}
\affiliation{Kellogg Radiation Laboratory, California Institute of Technology, Pasadena, California 91125, USA.}
\author{J. Ortiz}
\affiliation{Los Alamos National Laboratory, Los Alamos, NM 87544, USA}
\author{R. W. Pattie Jr.}
\affiliation{Department of Physics, North Carolina State University, Raleigh, North Carolina 27695, USA}
\author{D. G. Phillips II}
\affiliation{Department of Physics, North Carolina State University, Raleigh, North Carolina 27695, USA}
\author{J. Ramsey}
\affiliation{Los Alamos National Laboratory, Los Alamos, NM 87544, USA}
\author{D. J. Salvat}
\affiliation{Department of Physics,Indiana University, Indiana 47405-7105 USA}
\author{S. J. Seestrom}
\affiliation{Los Alamos National Laboratory, Los Alamos, NM 87544, USA}
\author{E. Shaw}
\affiliation{Los Alamos National Laboratory, Los Alamos, NM 87544, USA}
\author{Sky Sjue}
\affiliation{Los Alamos National Laboratory, Los Alamos, NM 87544, USA}
\author{W. E. Sondheim}
\affiliation{Los Alamos National Laboratory, Los Alamos, NM 87544, USA}
\author{B. VornDick}
\affiliation{Department of Physics, North Carolina State University, Raleigh, North Carolina 27695, USA}
\author{Z. Wang}
\affiliation{Los Alamos National Laboratory, Los Alamos, NM 87544, USA}
\author{T. L. Womack}
\affiliation{Los Alamos National Laboratory, Los Alamos, NM 87544, USA}
\author{A. R. Young}
\affiliation{Department of Physics, North Carolina State University, Raleigh, North Carolina 27695, USA}
\author{B. A. Zeck}
\affiliation{Department of Physics, North Carolina State University, Raleigh, North Carolina 27695, USA}


\date{May 31, 2013}

\begin{abstract}
The study of neutron cross sections for elements used 
as efficient ``absorbers'' of ultracold neutrons (UCN) is crucial for 
many precision experiments in nuclear and particle physics, cosmology and gravity.
In this context, ``absorption'' includes both the capture and upscattering of neutrons 
to the energies above the UCN energy region.
The available data, especially for hydrogen, do not agree between themselves 
or with the theory. 
In this report we describe measurements performed at the Los Alamos National Laboratory UCN facility  of the UCN upscattering cross sections 
for vanadium and for hydrogen in CH$_2$ using simultaneous measurements 
of the radiative capture cross sections for these elements. 
We measured $\sigma_{up}=1972\pm130$ b for hydrogen in CH$_2$, which is below theoretical expectations, and $\sigma_{up} < 25\pm9$ b for vanadium, in agreement with the expectation for the neutron heating by thermal excitations in solids.

\end{abstract}
\vspace{1pc}
\maketitle

Several recent reviews highlight the application of ultracold neutrons 
in past and ongoing experiments for searching for the neutron electric dipole moment 
\cite{Lamor09}, for precision measurement of the neutron lifetime \cite{Wiet011}, 
for neutron $\beta$-decay correlations \cite{Abele08} and for gravitational quantum states \cite{Nesv010},
as well as for solving new issues in cosmology \cite{Dubb011}. Ultracold neutrons greatly increase 
the sensitivity of experiments because they can be confined in the apparatus for 
times comparable with the neutron lifetime owing to the action of the `optical' Fermi potential, $U_F$, \\
\begin{equation}
U_F = \frac {2\pi \hbar^{2}} {m} {\rm Nb},
\end{equation}
where m is the neutron mass, N is the number density of the material, and b is the neutron scattering length.

Materials with high positive potential, $U_{F} \simeq 250$ neV, are used for trapping UCN 
because neutrons with kinetic energy lower than that value 
are reflected from walls at all angles of incidence. The velocity of such neutrons is below 7 m/s. 
On the other hand, materials made from elements with a small negative optical potential, like hydrogen, 
vanadium, titanium serve for removal of UCN through 
absorption and/or upscattering to higher energy. The efficiency of this removal is important for many 
measurements and this continues to motivate the study of such materials. Vanadium and hydrogen 
are perfect neutron incoherent scatterers, and both have small negative, practically non-reflective
potential $U_{F} \simeq -7.1$ neV.
For these elements, the theory of upscattering in one-phonon incoherent approximation is usually applied in the theory of upscattering, 
and the UCN isotropic differential upscattering cross section for cubic lattices can be calculated 
by a simple formula \cite{Gur68}:
\begin{equation}
\frac{d\sigma_{up}}{dE} = \sigma_{\rm b} \sqrt\frac {E}{E_{i}} \left (e^{E/kT}-1\right )^{-1}\frac {g(E)} {\rm A }e^{-2W},
\end{equation} 
where $\sigma_{\rm b} = 4\pi {\rm b}^2$ is the cross section for bound nuclei with the mass number ${\rm A }$, $E_{i}$ 
is the initial UCN energy, $E$ is the energy after upscattering, $g(E)$ is the phonon density of states 
and the last exponent is the Debye-Waller factor which is often measured, although it can be calculated as well. 
The \textsc{mcnp}  code \cite{Mcnp5} with the thermal data kernels for the scattering law $S(\alpha, \beta)$ 
($\alpha$ and $\beta$ are the reduced momentum and energy transfers) calculates 
the neutron inelastic scattering cross section corrected for multi-phonon contributions to Eq.(2) 
when applicable. The cross sections for several neutron moderators have been calculated using \textsc{mcnp}  by \cite{Cul03}. 
For polyethylene, at the minimum evaluated energy of 0.01 meV (v=43.7 m/s) the upscattering cross section is $\sigma_{up}=320$ b and
the 1/v law applies below $\simeq$ 0.1 meV. \\

Total neutron cross sections for vanadium have been measured over the 
velocity range from 4.4 m/s to 17.7 m/s in \cite{Polo83}. The inferred capture cross section 
was analyzed together with other thermal data and was shown to follow the expected 1/v law 
up to thermal velocities, 2200 m/s. With this result and with the thermal value 
$\sigma_a = 4.92\pm 0.05$ barn \cite{Mug06} for the isotope $^{51}$V, we write 
$\sigma_a\sqrt E = 0.808\pm 0.008$ b $eV^{1/2}$. 
The vanadium upscattering cross section $\sigma_{up}$ has not been measured but has been estimated 
\cite{Dilg74} in the one-phonon approximation to follow the 1/v law below $\simeq 0.1$ meV with 
the value $\sigma_{up}\sqrt E = 0.012\pm 0.002$ b $eV^{1/2}$, giving 
the relative contribution of the upscattering cross section 
at the level of 1.5\%. The capture cross section for $^{51}$V is large at UCN energies, 2700 b for UCN at 4 m/s 
velocity. 
For hydrogen the relationship between absorption and upscattering
is reversed: the average UCN upscattering cross section, 
according to the measurement \cite{Zhuk93}, is $3745 \pm370$ b while the thermal cross section 
value $\sigma_a = 0.3326\pm 0.0007$ \cite{Mug06} 
and the 1/v law leads to $\sigma_a {\rm (4\, m/s)}= 182.9$ b. 
Lower upscattering cross sections values for hydrogen in polyethylene
have been obtained 
from the recent UCN time-of-flight measurements \cite{Pok011}, where the $\sigma_{up} (E)$ was 
shown to follow the 1/v law with the value 
$\sigma_{up} {\rm (4\, m/s)}= 2053\pm 40$ b, at 4
 m/s .
The reasons for the disagreement between the results of measurements \cite{Zhuk93} 
and \cite{Pok011} remain unclear. \\

Measurements of the UCN upscattering and capture cross sections 
have been performed using the Los Alamos National Laboratory solid-deuterium 
ultracold neutron source 
driven by an 800-MeV, 5 $\mu$A average proton beam provided by the Los Alamos Neutron Science Center( LANSCE) linear accelerator. 
The source is described in detail in a recent publication by Saunders et al. \cite{Andy013}. 
The UCN density in the stainless steel guide after the source exit window was 
$\simeq 1$ UCN/cm$^3$. The UCN velocity spectrum 
was cut off by 
the $U_{F}=189$ neV potential. Our Monte Carlo modeling of the UCN transport 
in the source and guides \cite{Andy013} 
suggests a velocity spectrum $n(v)\sim v^{3}dv$ below cutoff, rather then a Maxwellian 
distribution $v^{2}dv$ 
because of absorption in the thick solid deuterium source.
This number was lower than typical because these measurements were made with non-ideal source conditions,
low volume and high para-deuterium contamination.
The cross sections obtained in this experiment, which are 
averaged over the UCN velocity spectrum, corresponds
to the average velocity $v_{av}\simeq 4$ m/s (the UCN energy $\simeq$80 neV). 
Upscattering and capture have been measured simultaneously.
The beam intensity during vanadium and hydrogen runs was monitored by our standard UCN monitor 
\cite{Andy013}. 
The gamma rays from neutron radiative capture in hydrogen and vanadium were measured in a 100\% 
efficient (relative to a 7.6 cm \diameter by 7.6 cm thick right circular cylinder of  NaIr) high purity germanium detector (HPGe).
The neutron detector forthe  upscattering measurement was a 30 cm long drift tubes \cite{Chris09} 
with 1.8 bar of 3He and 1 bar of CF$_4$. Both detectors were placed outside of the 7.5 cm diameter 
UCN guide upstream of the target location. The thicknesses of the targets, 
0.25 mm of vanadium, and 1.6 mm of polyethylene,
were chosen to absorb all neutrons that enter 
the materials. The target foils were backed by a nickel foil ($U_{F}({\rm Ni}) \simeq 250$ neV) 
between them and the end of a UCN 
guide, reflecting any transmitted UCN neutrons back into the sample. The weak 
reflections from the 
vanadium and polyethylene surfaces are nearly equal and negligible for UCN with  perpendicular energies
higher than $\simeq 5$ neV. \\

In Fig. 1 pulse height spectra from the neutron detector for the 
two targets are compared to a run with no target. The 
upscatter rate with the polyethylene target is much larger than with the vanadium 
target. Both targets produce upscatter neutron rates significantly above the 
background, which was mostly due to cosmogenic thermal background. The 
background associated with the proton beam was eliminated using time gates 
on the analog to digital converters, to reject events during the beam pulses. 
The relative neutron rates were calculated by integrating the spectra shown 
in Fig. 2 and subtracting the background measured with an empty target. 
The efficiencies of the $^3$H  detector for neutron spectra up scattered by CH$_2$ and V 
have been modeled using spectral data of \cite{Chris013} and found to be equal with the accuracy of 2\%.

In the measurement with the HPGE detector a large number of $\gamma$-ray lines were 
observed with both the vanadium and the polyethylene foils. With the polyethylene 
foil many of the lines were due to up scattered neutrons capturing in the materials of 
the detector. With the vanadium foil there are a large number of capture lines 
extending up to ~8 MeV.
Two $\gamma$-ray lines have been used in the current analysis, the 2.225 MeV n-p capture 
line and the 1.434 MeV $\gamma$ ray emitted in the decay of $^{52}$V. The HPeE photopeak 
efficiencies for them in our geometry have been modeled by the \textsc{mcnp5}  code with 
the result $\epsilon (1.434)$ = 1.37 $\epsilon (2.225)$. 
The $\gamma$ rate from 
the decay of the isotope $^{52}$V produced by neutron capture builds up to equilibrium with a time 
constant given by the lifetime of $^{52}$V of 5.4 minutes. Equilibrium was ensured by waiting 
for two lifetimes with the beam on target before starting measurements with the 
vanadium target. 
The resulting spectra are shown in Fig. 2. Rates were extracted by subtracting 
the background measured using the alternate target. The empty runs were not useful 
for these because of the large increase in background due to Compton events 
from other capture induced processes in the detector.\\

\begin{table}[ht]
\caption{Neutron $r_n$ and gamma ray $r_a$ count per 300 s, their ratios R(V/H) and the obtained UCN
scattering cross sections $\sigma_{n}$ for the Vanadium target and Hydrogen in the polyethylene
target.
The gamma ray ratio has been corrected for the relative efficiency for the
vanadium(1.44 MeV) and hydrogen capture (2.22 MeV) lines. The absorption cross sections
$\sigma_a$, used in the analysis, are obtained by applying the 1/v law to
thermal data from \cite{Mug06}. }
\label{tab:tb1}
\begin{tabular} {||c|c|c|c|c||}\hline\hline
\rule{0pt}{4ex}Targets and R(V/H) & $r_n$ &$r_a$ &$\sigma_{n}(4\;{\mathrm m/s}),\; {\mathrm b}$
&$\sigma_a(4\;{\mathrm m/s}),\; {\mathrm b}$ \\
\hline
\rule{0pt}{4ex} Vanadium (V) &$19\pm 5$ &$10590\pm 11$ &25$\pm$ 9 & 2706$\pm$ 0.27 \\
Polyethylene (H) &$1869\pm 22$ & $871\pm 40$ &2062$\pm$ 135 & 182.9$\pm$ 0.4 \\
\hline
$R_n(V/H)$ & $0.010\pm 0.003$ & & & \\
$R_a(V/H)$ & & $12.16\pm 0.7$& & \\
\hline
\hline
\end{tabular}
\end{table}

The data were analyzed as follows. Because all the incident UCN on V and CH$_2$ are 
either captured or scattered, the relative contributions of these processes are 
given by the ratios 
$\sigma_a/ \sigma_t$ and $\sigma_n/ \sigma_t$, where the total cross section 
$\sigma_t$ is the sum $\sigma_t = \sigma_n + \sigma_a$. The counting rates 
of the capture 
and neutron detectors, $r_a$ and $r_n$, corrected for efficiencies 
in the V and H targets are 
\begin{equation}
r_a = \frac{\sigma_a}{\sigma_t} \Phi S \epsilon_a , \;\; 
r_n = \frac{\sigma_n}{\sigma_t} \Phi S \epsilon_n ,
\end{equation}
where $\Phi$ is the neutron fluence (the same for both targets after normalizing on the UCN monitor), 
$S$ is the target area (the same for both targets), 
$\epsilon_a$ and $\epsilon_n$ are the corresponding detectors efficiencies (including 
the solid angles) and we omitted, for simplicity, the indices of targets.

Introducing the ratios between vanadium and hydrogen count rates 
\begin{equation}
R_a \equiv \frac {r_a(V)}{r_a(H)} = \frac{\sigma_a(V)}{\sigma_t(V)} \frac{\sigma_t(H)}{\sigma_a(H)} ,
\end{equation}
and
\begin{equation}
R_n \equiv \frac {r_n(V)}{r_n(H)} = \frac{\sigma_n(V)}{\sigma_t(V)} \frac{\sigma_t(H)}{\sigma_n(H)} ,
\end{equation}
we obtain the relationship between the measured V and H cross sections. 
\begin{equation}
\frac{\sigma_a(V)}{\sigma_n(V)} = \frac {R_a}{R_n} \frac{\sigma_a(H)}{\sigma_n(H)} .
\end{equation}
To obtain relationships between $\sigma_a$ and $\sigma_n$ for each target separately 
one has to estimate the quantities $(1-R_n)$ and $(R_a -1)$ making use of the inequality 
$\sigma_a(V)\gg \sigma_n(V)$. The result for hydrogen is
\begin{equation}
\frac{\sigma_a(H)}{\sigma_n(H)} = \frac {(1-R_n)}{(R_a-1)},
\end{equation}
and, after substitution it into Eq. (6), the result for vanadium is
\begin{equation}
\frac{\sigma_a(V)}{\sigma_n(V)} = \frac {R_a}{R_n}\frac {(1-R_n)}{(R_a-1)}.
\end{equation}

The experimental results obtained from this analysis are given in Table I. 
From these data the UCN upscattering cross section on hydrogen, 
taken as the difference between the measured total scattering cross section 
$\sigma_n$ and the known (90 b) 
elastic $\sigma_{el}$ cross section, is $\sigma_{up}(4\;{\mathrm m/s})= 1972\pm 130$ b. 
Our measurements support recent TOF measurements of 
Pokotilovski et al. \cite{Pok011}, but disagree with the measurements of Zhukov et al. \cite{Zhuk93}, 
which served the UCN community for a long time as the only experimental data on 
upscattering in CH$_2$. It should be noted that our \textsc{mcnp}  
calculations \cite{Cul03} of $\sigma_{up}$ for CH$_2$ agree with \cite{Zhuk93} in disagreement with our
experimental result. 
Our result for vanadium is the first 
experimental confirmation of the extremely low UCN upscattering in V and, in this case, 
agrees with the theory of UCN heating by phonons in solids. \\

{\bf Acknowledgment} This work was performed under the auspices of the U.S. 
Department of Energy under Contract DE-AC52-06NA25396. Author DJS is supported by the DOE Office of Science Graduate Fellowship Program (DOE SCGF), made possible in part by the American Recovery and Reinvestment Act of 2009, administered by ORISE-ORAU under contract no. DE-AC05-06OR23100.

\begin{figure}[h] 
\centering
\includegraphics[width=7.0in,angle=0]{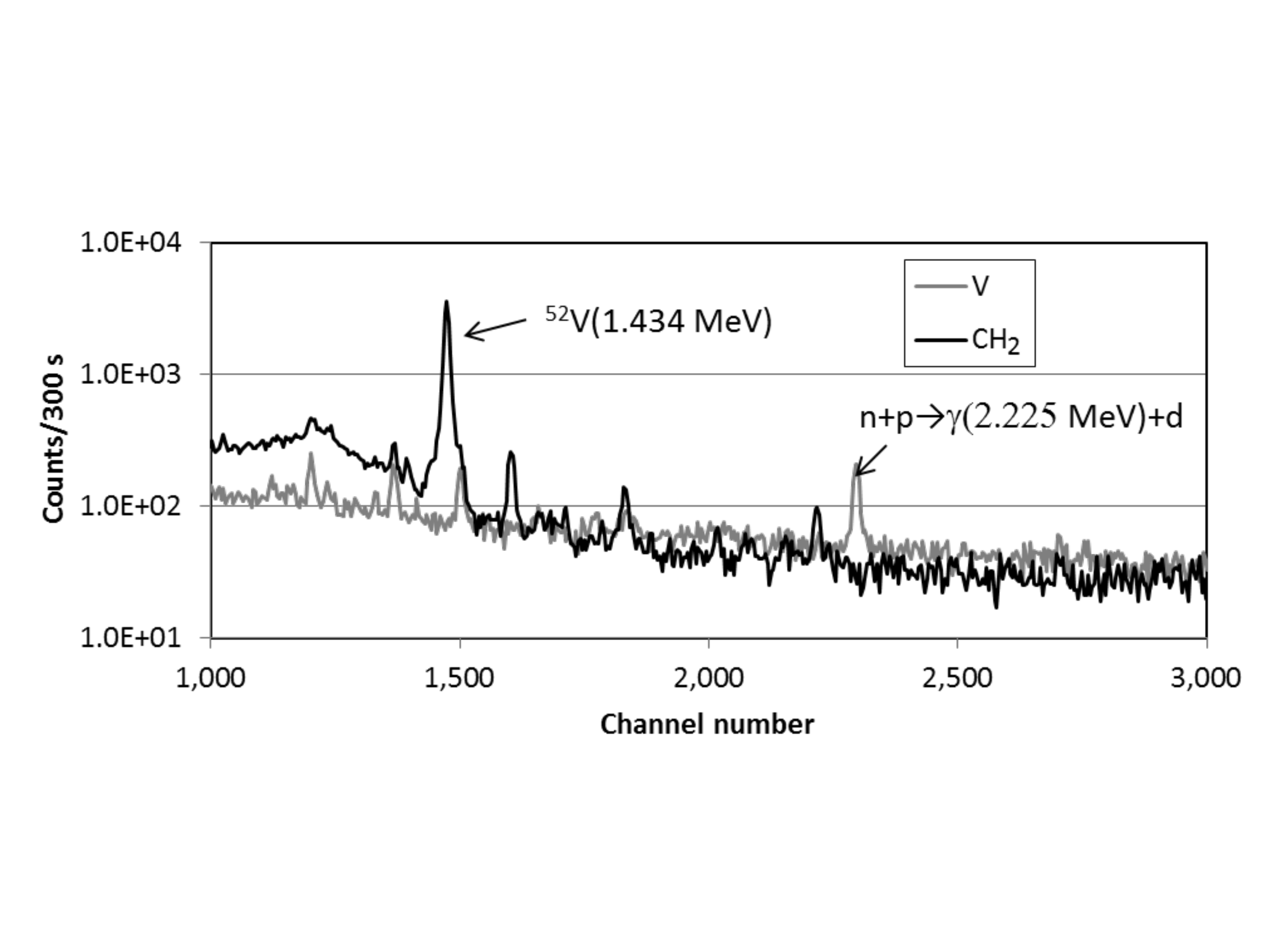}
\caption{Pulse height spectra from the 3He detector for the polyethylene, vanadium and no target. 
The shoulder on the large pulse height peak and the low pulse height peak are due to wall effects \
in the detector.}
\label{Fig1}
\end{figure}

\begin{figure}[b] 
\centering
\includegraphics[width=7.0in,angle=0]{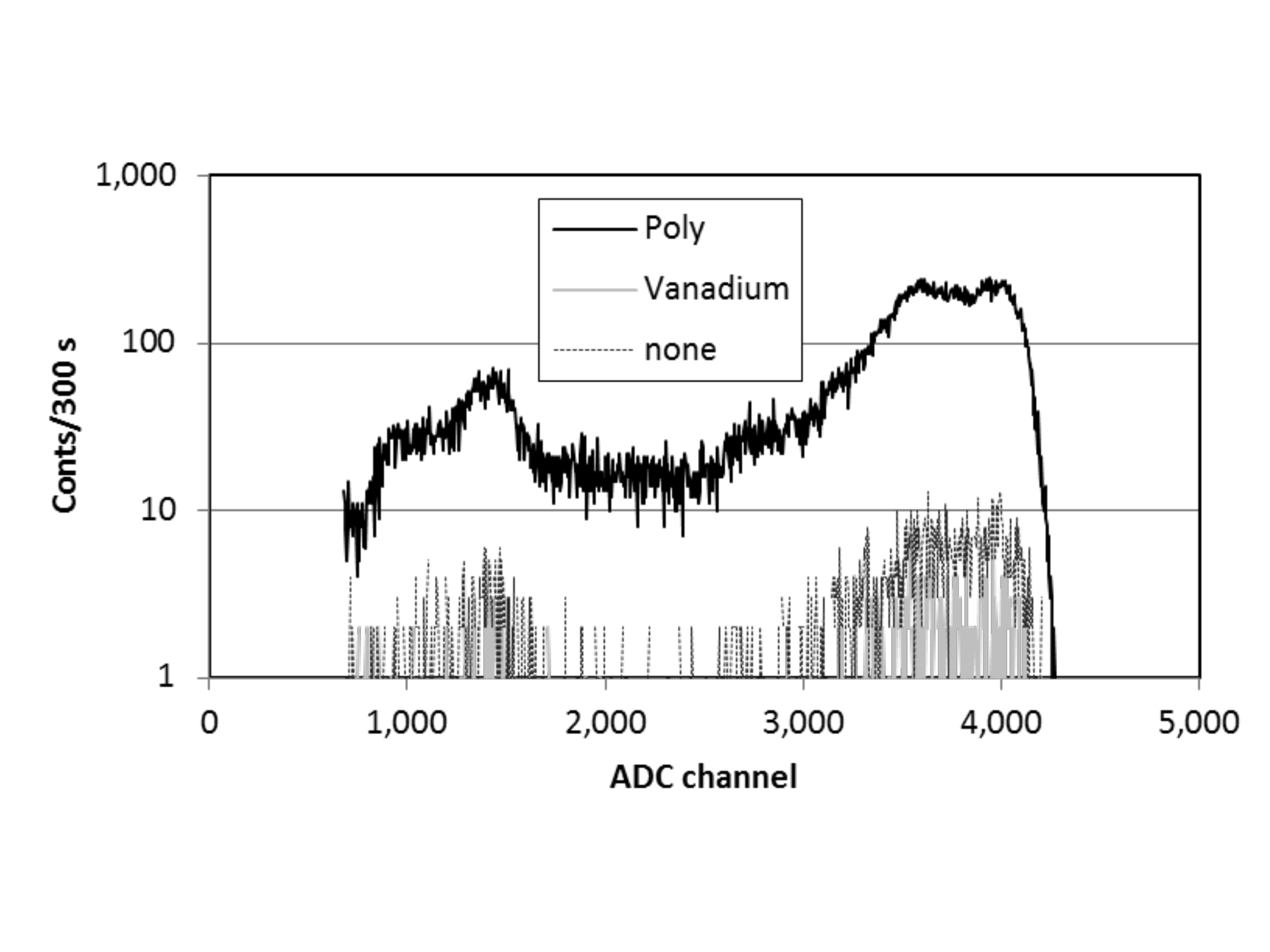}
\caption{ Gamma ray spectra obtained with UCN. The gamma line from n-p capture with 
the polyethylene target and the gamma line from the product target $^{52}$V 
are labeled.} 
\label{Fig2}
\end{figure}

\end{document}